\documentclass{ws-procs9x6}

\usepackage{graphicx}

\begin{document}

\title{EXPOSING THE DRESSED QUARK'S MASS}

\author{H.\,L.\,L.~ROBERTS,$^{1,2}$ L.~CHANG,$^{3}$ I.\,C.~CLO\"ET,$^{4}$ and C.\,D.~ROBERTS$^{1,2,5}$}

\address{$^1$Physics Division, Argonne National Laboratory\\ Argonne, Illinois 60439, USA
\\
$^2$Institut f\"ur Kernphysik,
Forschungszentrum J\"ulich\\ D-52425 J\"ulich, Germany\\
$^3$Institute of Applied Physics and Computational Mathematics\\ Beijing 100094, China\\
$^4$Department of Physics, University of Washington\\ Seattle WA 98195, USA\\
$^5$Department of Physics, Peking University\\ Beijing 100871, China}

\begin{abstract}
This snapshot of recent progress in hadron physics made in connection with QCD's Dyson-Schwinger equations includes: a perspective on confinement and dynamical chiral symmetry breaking (DCSB); a pr\'ecis on the physics of in-hadron condensates; results on the hadron spectrum, including dressed-quark-core masses for the nucleon and $\Delta$, their first radial excitations, and the parity-partners of these states; an illustration of the impact of DCSB on the electromagnetic pion form factor, thereby exemplifying how data can be used to chart the momentum-dependence of the dressed-quark mass function; and a prediction that $F_1^{p,d}/F_1^{p,u}$ passes through zero at $Q^2\approx 5m_N^2$ owing to the presence of nonpointlike scalar and axial-vector diquark correlations in the nucleon.
\end{abstract}


\bodymatter

\vspace*{4ex}

%\section{Mass of a Dressed-quark}
%
\hspace*{-\parindent}\textbf{1.~Confinement, DCSB and in-hadron condensates}.
Quantum chromodynamics is a theory whose elementary excitations are not those degrees-of-freedom readily accessible via experiment; i.e., whose elementary excitations are confined.  Moreover, less-than 2\% of a nucleon's mass can be attributed to the so-called current-quark masses that appear in QCD's Lagrangian, with the remainder being generated through dynamical chiral symmetry breaking (DCSB).\cite{Flambaum:2005kc}  Neither confinement nor DCSB is apparent in QCD's Lagrangian and yet they play the dominant role in determining the observable characteristics of real-world QCD.  The physics of hadrons is ruled by such \emph{emergent phenomena}, which can only be elucidated via nonperturbative methods in quantum field theory.  This is both the greatest novelty and the greatest challenge within the Standard Model.

\begin{figure}[t]
\includegraphics[clip,width=0.5\textwidth]{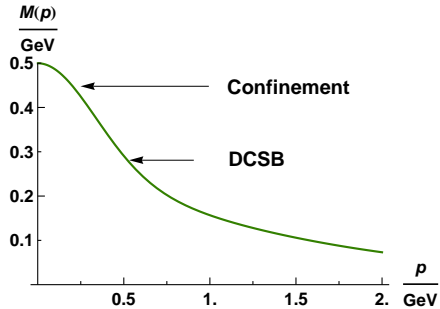}
\vspace*{-23ex}

\rightline{\parbox{0.45\textwidth}{
\caption{\label{MpFig} Dressed-quark mass function used, e.g., in an extensive study of nucleon electromagnetic form factors.\protect\cite{Cloet:2008re}  Dynamical chiral symmetry breaking is evident in the enhancement around $p=0.5\,$GeV; and confinement is signalled by the inflexion point at around $p=0.25$\,GeV.}}}
\end{figure}

Both confinement and DCSB can be seen in the dressed-quark mass function, illustrated by Fig.\,\ref{MpFig}.  The behaviour of $M(p)$ can be calculated in perturbation theory for $p\gtrsim 1.5\,$GeV.  %This is a textbook exercise.
However, the evolution of $M(p)$ is essentially nonperturbative for infrared momenta.  The existence of DCSB is signalled by the rapid increase in magnitude as $p$ decreases from $1 \to 0.5\,$GeV.  This longstanding prediction of Dyson-Schwinger equation (DSE) studies in QCD, see e.g. Refs.\,[\refcite{Bhagwat:2003vw,Bhagwat:2006tu}], is confirmed by numerical simulations of lattice-QCD.\cite{Bowman:2005vx}  The reality of DCSB means that the Higgs mechanism is largely irrelevant to the bulk of normal matter in the universe.  %Instead the single most important mass generating mechanism for light-quark hadrons is the strong interaction effect of DCSB.%; e.g., one can identify it as being responsible for 98\% of a proton's mass.

In analysing large momentum transfer processes it is common to employ a light-front formulation of QCD.  This has merit but there appears to be a serious drawback; viz., DCSB has not yet been realised on the light-front.  However, progress has recently been made toward resolving this conundrum,\cite{Brodsky:2010xf} by arguing for a shift in paradigm so that, in particular, DCSB is understood as being expressed in properties of hadrons rather than of the vacuum.  It is contended therein that: the measurable impact of all condensates is entirely expressed in the properties of QCD's asymptotically realisable states; there is no evidence that the so-called vacuum condensates are anything more than mass-dimensioned parameters in one or another theoretical truncation scheme; and condensates do not describe measurable spacetime-independent configurations of QCD's elementary degrees-of-freedom in a hadron-less ground state.  This position assumes confinement, from which follows quark-hadron duality and hence that all observable consequences of QCD can, in principle, be computed using a hadronic basis.

\begin{figure}[t]
\includegraphics[clip,width=0.48\textwidth]{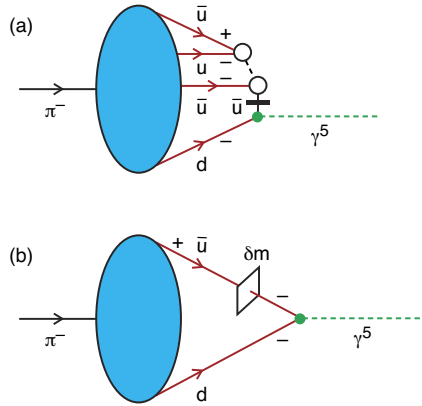}
\vspace*{-35ex}

\rightline{\parbox{0.45\textwidth}{\caption{\label{instantaneous}
Light-front contributions to $\rho_\pi=-\langle 0| \bar q \gamma_5 q |\pi\rangle$.
\emph{Upper panel} -- A non-valence piece of the meson's light-front wavefunction, whose contribution to $\rho_\pi$ is mediated by the light-front instantaneous quark propagator (vertical crossed-line).  The ``$\pm$'' denote parton helicity.
\emph{Lower panel} -- There are infinitely many such diagrams, which can introduce chiral symmetry breaking in the light-front wavefunction in the absence of a current-quark mass.
(The case of $f_\pi$ is analogous.)}}}\vspace*{5ex}
\end{figure}

From this perspective a possible solution to the problem of DCSB in the light-front formulation is illustrated in Fig.\,\ref{instantaneous}.  The light-front-instantaneous quark propagator can mediate a contribution from higher Fock state components to the matrix elements that define the pion's pseudoscalar ($\rho_\pi$) and pseudovector ($f_\pi$) decay constants.  Such diagrams connect dynamically-generated chiral-symmetry breaking components of the meson's light-front wavefunction to these matrix elements.  There are infinitely many contributions of this type and they do not depend sensitively on the current-quark mass in the neighborhood of the chiral limit.  This leads to a conjecture that DCSB in the light-front formulation is expressed via in-hadron condensates and is connected with sea-quarks derived from higher Fock states.

Confinement is signalled in Fig.\,\ref{MpFig} by the inflexion point possessed by $M(p)$ at $p\approx 0.25$\,GeV, a relationship which is explained, e.g., in Refs.\,[\protect\refcite{Krein:1990sf,Roberts:2007ji,Chang:2010jq}].  Confinement and DCSB are intimately connected in QCD, and it is natural to ask whether the connection is accidental or causal.  There are models with DCSB but not confinement, however, a model with confinement but lacking DCSB has not yet been identified (see, e.g., Secs.\,2.1 and 2.2 of Ref.\,[\refcite{Roberts:2007jh}]).  This leads to a conjecture that DCSB is a necessary consequence of confinement.  It is notable that there are numerous models and theories which exhibit both confinement and DCSB, and possess an external control parameter such that deconfinement and chiral symmetry restoration occur simultaneously at some critical value of this parameter; e.g., quantum electrodynamics in three dimensions with $N_f$ electrons \cite{Bashir:2008fk,Bashir:2009fv}.

These observations highlight that confinement can be related to the analytic properties of QCD's Schwinger functions,\cite{Krein:1990sf,Roberts:2007ji,Chang:2010jq} which are often called Euclidean-space Green functions.  From this standpoint the question of light-quark confinement can be translated into the challenge of charting the infrared behavior of QCD's \emph{universal} $\beta$-function.  This is a well-posed problem whose solution is an elemental goal of modern hadron physics and which can be addressed in any framework enabling the nonperturbative evaluation of renormalisation constants.

\smallskip

\hspace*{-\parindent}\textbf{2.~DCSB and the hadron spectrum}.
Through the gap and Bethe-Salpeter equations (BSEs) the pointwise behaviour of the $\beta$-function determines the pattern of chiral symmetry breaking; e.g., the behaviour in Fig.\,\ref{MpFig}.  Moreover, the fact that these and other DSEs connect the $\beta$-function to experimental observables entails, e.g., that comparison between computations and observations of the hadron mass spectrum, and hadron elastic and transition form factors, can be used to chart the $\beta$-function's long-range behaviour.\cite{Aznauryan:2009da}  In order to realise this goal, a nonperturbative symmetry-preserving DSE truncation is necessary.  Steady quantitative progress can be made with a scheme that is systematically improvable \cite{Munczek:1994zz,Bender:1996bb}.  In fact, the mere existence of such a scheme has enabled the proof of exact nonperturbative results in QCD.\cite{Holl:2004fr,Holl:2005vu,Bhagwat:2007ha}

On the other hand, significant qualitative advances in understanding the essence of QCD could be made with symmetry-preserving kernel \emph{Ans\"atze} that express important additional nonperturbative effects, which are impossible to capture in any finite sum of contributions. Such an approach is now available\cite{Chang:2009zb}.  It begins with a novel form for the axial-vector BSE, which is valid when the quark-gluon vertex is fully dressed.  Therefrom, a Ward-Takahashi identity for the Bethe-Salpeter kernel is derived and solved for a class of dressed-quark-gluon-vertex models.  The solution yields a symmetry-preserving closed system of gap and vertex equations.  As the analysis can readily be extended to the vector equation, a comparison is possible between the responses of parity-partners in the meson spectrum to nonperturbatively dressing the quark-gluon vertex.  The results indicate that: spin-orbit splitting in the ground-state meson spectrum is dramatically enhanced by DCSB;\cite{Chang:2009zb} DCSB generates a large anomalous chromomagnetic moment term as an essential part of the dressed-quark-gluon vertex;\cite{Chang:2010jq} and, importantly, owing to the symmetry-preserving nature of the truncation procedure, the $M(p)$-driven vertex corrections alter the Bethe-Salpeter kernel in such a way as to leave ground-state pseudoscalar and vector meson masses almost unchanged.\cite{Chang:2010jq}

Motivated by this ongoing work, we employed the DSE model formulated in Ref.\,[\refcite{GutierrezGuerrero:2010md}] to compute a spectrum of ground-state $u\,$\&$\,d$-quark mesons, with the results presented in Table\,\ref{masses}.  The first row lists results obtained in rainbow-ladder truncation.  Of course, given the symmetry preserving nature of the truncation, in the chiral-limit $m_\pi=0$ and $m_\sigma = 2 M$ (which follows from Eq.\,(14) in Ref.\,[\refcite{GutierrezGuerrero:2010md}]), where $M=0.40\,$GeV is chiral-limit value of the model's momentum-independent dressed-quark mass.  The second row lists values obtained with spin-orbit repulsion added to the scalar and axial-vector channels, the strength of which is described by a single parameter, $g_{\rm SO}=0.32$, fixed so as to yield $m_{a_1}=1.23\,$GeV.  (We emphasise that introducing $g_{\rm SO}\neq 1$ is an expedient that mimics those effects of a \emph{momentum-dependent} mass function exposed and quantified in Refs.\,[\refcite{Chang:2010jq,Chang:2009zb}].)  It will be noted that $m_\sigma$ increases to a value which matches an estimate for the $\bar q q$-component of this meson obtained using unitarised chiral perturbation theory.\cite{Pelaez:2006nj}

%pi sigma rho a1 qqS qqP qqA qqV N N- N* D D- D*
\begin{table}[t]
\tbl{\label{masses}
Meson and diquark masses (in GeV) computed using a contact-interaction DSE kernel,\protect\cite{GutierrezGuerrero:2010md} which produces a momentum-independent dressed-quark mass $M=0.41\,$GeV from a current-quark mass of $m=8\,$MeV.  ``RL'' denotes rainbow-ladder truncation.\protect\cite{Bender:1996bb}  Row-2 is obtained by augmenting the RL kernel with spin-orbit repulsion, as described in the text.  For reference, experimental masses are:\protect\cite{Amsler:2008zzb} $m_\pi=0.140$, $m_\sigma = 0.4-1.2$, $m_\rho=0.775$, $m_{a_1}=1.243$.  NB.\ We implement isospin symmetry so, e.g., $m_\omega=m_\rho$, $m_{f_1}=m_{a_1}$, etc.}
{\begin{tabular}{@{}lccccccccc@{}}\toprule
   & $m_\pi$ & $m_\sigma$ & $m_\rho$ & $m_{a_1}$ & $m_{0^+}^{qq}$ & $m_{0^-}^{qq}$ & $m_{1^+}^{qq}$ & $m_{1^-}^{qq}$ & \\[1ex]
RL & 0.141 & 0.825 & 0.798 & 1.073 & 0.723 & 0.975 & 1.006 & 1.173 \\
RL $+ \,g_{\rm SO}$ & 0.141 & 1.079 & 0.798 & 1.243 & 0.723 & 1.189 & 1.006 & 1.323
\end{tabular}}
\end{table}

In quantum field theory an analogue for baryons of the BSE for mesons is a Poincar\'e covariant Faddeev equation that sums all possible exchanges and interactions that can take place between three dressed-quarks.  A tractable Faddeev equation\cite{Cahill:1988dx} is founded on the observation that an interaction which describes colour-singlet mesons also generates quark-quark (diquark) correlations in the colour-$\bar 3$ (antitriplet) channel.\cite{Cahill:1987qr}  In QCD these correlations are \emph{nonpointlike}:\cite{Cloet:2008re} $r_{qq} \approx 0.8\,$fm.

Within the model of Ref.\,[\refcite{GutierrezGuerrero:2010md}] we have formulated the Faddeev equation for $u\,$\&$\,d$-quark octet and decuplet baryons.   The kernel involves the propagation of a dressed-quark, which originates at the breakup of one diquark and ends at the formation of another (see Fig.\,1 of Ref.\,[\refcite{Cloet:2008re}]).  The equations are particularly simple if one uses an additional truncation:
%\begin{equation}
%\label{staticT}
$
%S_{\rm exchange} \to \frac{g_B}{M}\,,
S_{\rm exchange} \to (g_B/M)\,,
$
%\end{equation}
based on the so-called \emph{static approximation},\cite{Buck:1992wz}
but here with a parameter $g_B$ introduced to restore attraction that is lost through this expedient.  Now the Faddeev equation for the ground-state $\Delta$ reduces to the following eigenvalue problem for $m_\Delta$:
\begin{equation}
\label{DeltaFE}
2 \pi^2 = \frac{g_B}{M} \frac{1}{m_{1^+}^2}\, \int_0^1 \!\! d\alpha\; 4\!\int_q^\Lambda \Gamma_{1^+}^2
\frac{(m_{1^+}^2 +  (1-\alpha)^2 m_\Delta^2) (\alpha \, m_\Delta+M)}{(q^2 + (1 - \alpha) M^2 + \alpha m_{1^+}^2 - \alpha (1 - \alpha) m_\Delta^2)^2},
\end{equation}
where $\Gamma_{1^+}$ is the canonically normalised Bethe-Salpeter amplitude for the $1^+$-diquark and $\int_q^\Lambda := \int d^4 q/(4\pi^2)$, regularised as explained in Ref.\,[\refcite{GutierrezGuerrero:2010md}].  The $\Delta$ is particularly simple because it only involves an axial-vector diquark.  It could conceivably also contain a vector diquark component.  However, that has conflicting parity and hence can only appear in combination with material orbital angular momentum between the quark and vector diquark, which is impossible in our model and strongly suppressed in Nature.  The nucleon equation is more complicated than Eq.\,(\ref{DeltaFE}) because one must combine scalar and axial-vector diquark components but it remains quite simple: in general, a $5\times 5$-matrix eigenvalue problem for $m_N$.

\begin{table}[t]
\tbl{\label{massesB}
\emph{Row-1}: Dressed-quark-core masses for nucleon and $\Delta$, their first radial excitations (denoted by $\ast$), and the parity-partners of these states, computed with $g_N=1.33$, $g_\Delta=1.58$.  \emph{Row-2}: Bare-masses inferred from a coupled-channels analysis  at the Excited Baryon Analysis Center (EBAC).\protect\cite{Suzuki:2009nj}  EBAC's method does not provide a bare nucleon mass and ``{\ldots}'' indicates states not found in their analysis.}
{\begin{tabular}{@{}lcccccccc@{}}\toprule
& $m_N$ & $m_{N^\ast}$ & $m_{N \frac{1}{2}^{-}}$ & $m_{N^\ast \frac{1}{2}^{-}}$
& $m_\Delta$ & $m_{\Delta^\ast}$ & $m_{\Delta \frac{3}{2}^{-}}$ & $m_{\Delta^\ast \frac{3}{2}^{-}}$\\
This work & 1.05 & 1.73 & 1.89  & 2.10 & 1.33 & 1.85 & 2.01 & 2.18\\
EBAC      &      & 1.76 & 1.80  & \ldots & 1.39 & \ldots & 1.98 & \ldots %\\
%          &      &      & 1.88  &        &      &        &      &
\end{tabular}}
\end{table}

In Table\,\ref{massesB} we report preliminary results for the dressed-quark-core masses of the ground-state nucleon and $\Delta$, their first radial excitations, and the parity-partners of these states.  This is a first step toward the theory goals described in Ref.\,[\refcite{Aznauryan:2009da}].
The result $m_\Delta \approx 3 M$, which is a good approximation over a wide range of current-quark masses and interaction strengths, emphasises the simplicity of the $\Delta$ resonance.
Naturally, in quantum field theory as in quantum mechanics, the bound-state amplitude for a first radial excitation must possess a single zero.\cite{Holl:2004fr}  We implement that feature via the method employed in Ref.\,[\refcite{Volkov:1999yi}]; i.e., for instance, by introducing a factor $(1- d_{\rm r} q^2)$ on the right-hand-side in Eq.\,(\ref{DeltaFE}) with $d_{\rm r} = 1/[2 \Lambda_{\rm ir}^2]$ where $1/\Lambda_{\rm ir} = 0.83\,$fm is the model's confinement length-scale.\cite{GutierrezGuerrero:2010md}  We view the agreement between our dressed-quark-core masses and EBAC's bare masses as significant, since no attempt was made to ensure this.  Our results hint that $N(2090)\,S_{11}$ is the first radial excitation of $N(1535)\,S_{11}$.

\smallskip

\begin{figure}[t] % fig4
\includegraphics[clip,width=0.40\textwidth,angle=-90]{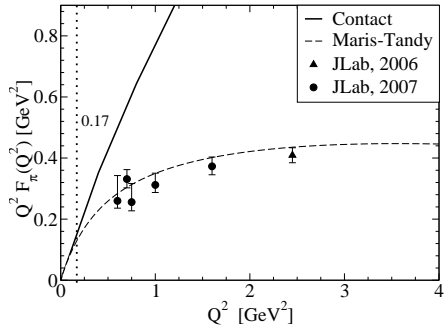}
\vspace*{-30ex}

\rightline{\parbox{0.45\textwidth}{
\caption{\label{figFpi} \emph{Solid curve}: Result obtained for $Q^2 F_{\pi}(Q^2)$
using a contact interaction.\protect\cite{GutierrezGuerrero:2010md}  \emph{Dashed curve}: DSE prediction \protect\cite{Maris:2000sk}, which employed a momentum-de\-pen\-dent renormalisation-group-im\-pro\-ved gluon exchange interaction. For $Q^2>0.17\,$GeV$^2\approx M^2$, marked by the vertical \emph{dotted line}, the contact interaction result for $F_{\pi}(Q^2)$ differs from that in Ref.\,[\protect\refcite{Maris:2000sk}] by more than 20\%.  The data are from Refs.\,[\protect\refcite{Horn:2006tm,Tadevosyan:2007yd}].}}}
\end{figure}

\hspace*{-\parindent}\textbf{3.~Elastic form factors}.
Once one has bound-state masses and Bethe-Salpeter or Faddeev amplitudes, the computation of hadron elastic and transition form factors can proceed.  The DSE calculations of $F^{\rm em}_\pi(Q^2)$ in Refs.\,[\refcite{Maris:1998hc,Maris:2000sk}] are an archetype ($Q^2$ is the squared-momentum-transfer).  The most systematic of these\cite{Maris:2000sk} predicted the measured form factor.\cite{Volmer:2000ek}  An elucidation of the sensitivity of $F^{\rm em}_\pi(Q^2)$ to the pointwise behaviour of the interaction between quarks is the main theme of Ref.\,[\refcite{GutierrezGuerrero:2010md}]; and in Fig.\,\ref{figFpi} we compare the form factor computed using a contact-interaction with the QCD-based DSE prediction\cite{Maris:2000sk} and contemporary experimental data.\cite{Volmer:2000ek,Horn:2006tm,Tadevosyan:2007yd}  Both the QCD-based result and that obtained from the momentum-in\-de\-pen\-dent interaction yield the same values for the pion's static properties. However, for $Q^2>0$ the form factor computed using $\sim 1/k^2$ vector boson exchange is immediately distinguishable empirically from that produced by a momentum-independent interaction.  Indeed, the figure shows that for $F_\pi^{\rm em}$, existing experiments can already distinguish between different possibilities for the quark-quark interaction.

The analysis of Ref.\,[\refcite{GutierrezGuerrero:2010md}] demonstrates that when a momentum-independent vector-exchange interaction is regularised in a symmetry-preserving manner, the results are directly comparable with experiment, computations based on well-defined and systematically-improvable truncations of QCD's DSEs\cite{Maris:2000sk}, and perturbative QCD.  However, a contact interaction, whilst capable of describing pion static properties well, generates a form factor whose evolution with $Q^2$ deviates markedly from experiment for $Q^2>0.17\,$GeV$^2\approx M^2$ and produces asymptotic power-law behaviour: $Q^2 F_\pi(Q^2)\approx\,$constant, in serious conflict with perturbative-QCD \cite{Farrar:1979aw,Efremov:1979qk,Lepage:1980fj}.

It is noteworthy that the contact interaction produces a momentum-independent dressed-quark mass function, in contrast to QCD-based DSE studies\cite{Bhagwat:2003vw,Bhagwat:2006tu,Roberts:2007ji} and lattice-QCD.\cite{Bowman:2005vx}  This is the origin of the marked discrepancy between the form factor it produces and extant experiment.  Hence Ref.\,[\refcite{GutierrezGuerrero:2010md}] highlights that form factor observables, measured at an upgraded Jefferson laboratory, e.g., are capable of mapping the running of the dressed-quark mass function.  Efforts are underway to establish the signals of the running mass in baryon elastic and transition form factors.  In this connection one statement can readily be made.  Faddeev and Bethe-Salpeter amplitudes are peaked at zero relative momentum.  Hence, as demonstrated explicitly in Ref.\,[\refcite{Roberts:1994hh}], the domain of greatest support in the impulse approximation calculation of elastic form factors is that with each quark absorbing: mesons, $Q/2$; and baryons, $Q/3$.  The dressed-quark mass, $M(p^2)$, is perturbative for $p^2>2\,$GeV$^2$.  It can be argued from these observations that pQCD behaviour will not be observed in meson form factors unless $Q^2>8\,$GeV$^2$ and not in baryon form factors unless $Q^2>18\,$GeV$^2$.

\begin{figure}[t]
%\vspace*{-7ex}

\includegraphics[clip,width=0.50\textwidth]{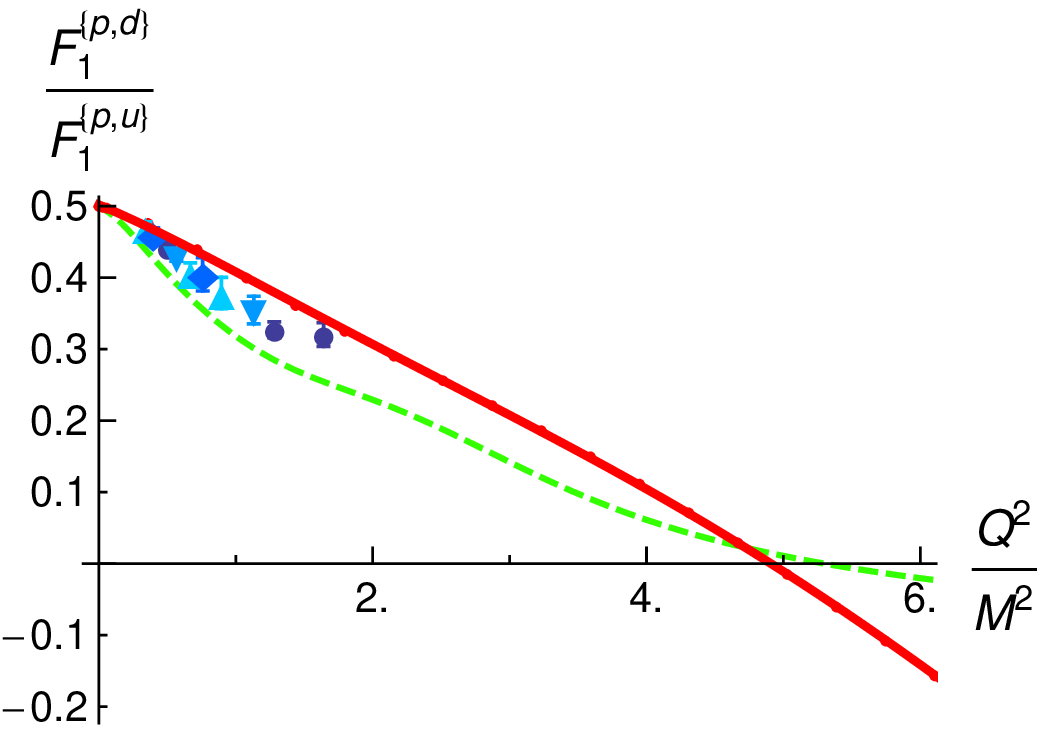}
\vspace*{-28ex}

\rightline{\parbox{0.45\textwidth}{\caption{\label{F1flavour}
Computed ratio of flavour-separated contributions to proton's Dirac form factor ($M =\,$nucleon mass) \emph{solid curve}.  Data, reconstructed from selected neutron electric form factor data using parametrisations of Ref.\,[\protect\refcite{Kelly:2004hm}] when necessary:\protect\cite{SRiordan}
Ref.\,[\protect\refcite{Glazier:2004ny}], \emph{up-triangles};
Ref.\,[\protect\refcite{Bermuth:2003qh}], \emph{diamonds};
Ref.\,[\protect\refcite{Plaster:2005cx}], \emph{circles};
and Ref.[\,\protect\refcite{Warren:2003ma}], \emph{down-triangles}.
\emph{Dashed-curve} -- This ratio computed from a recent parametrisation of nucleon form factor data.\protect\cite{Bradford:2006yz}.}}}
\end{figure}

Numerous nucleon properties have been computed via a more sophisticated version of the Faddeev equation than that used above to produce Table\,\ref{massesB}.\cite{Flambaum:2005kc,Cloet:2008re,Cloet:2008wg,Cloet:2008fw,Chang:2009ae}  Herein we draw only one example from a comprehensive analysis of nucleon electromagnetic form factors\cite{Cloet:2008re} that incorporates the momentum-dependent mass function illustrated in Fig.\,\ref{MpFig}.  In Fig.\,\ref{F1flavour} we depict a ratio of flavour-separated contributions to the proton's Dirac form factor.  (NB.\ When isospin is a good symmetry, $F_1^{n,u}/F_1^{n,d} = F_1^{p,d}/F_1^{p,u}$.)  The predicted $Q^2$-dependence owes to the presence of axial-vector diquark correlations in the nucleon.  It has been found\cite{Cloet:2008re} that the proton's singly-represented $d$-quark is more likely to be struck in association with an axial-vector diquark correlation than with a scalar, and form factor contributions involving an axial-vector diquark are soft.  On the other hand, the doubly-represented $u$-quark is predominantly linked with harder scalar-diquark contributions.  This produces a $d$-quark Dirac form factor which is softer than that of $u$-quark and a ratio $F_1^{p,d}/F_1^{p,u}$ that passes through zero.  The location of the zero depends on the relative probability of finding $1^+$ and $0^+$ diquarks in the proton.  The same physics explains the $x=1$ value of the $d_v(x)/u_v(x)$ ratio of valence-quark distribution functions in the proton.\cite{Holt:2010vj}

\smallskip

\hspace*{-\parindent}\textbf{4.~Epilogue}.
Herein we have exemplified the dramatic impact that DCSB has upon observables.  The behaviour of the dressed-quark mass function heralds DCSB; and the momentum dependence manifest in Fig.\,\ref{MpFig} is an essentially quantum field theoretical effect.  Exposing and elucidating its consequences therefore requires a nonperturbative and symmetry-preserving approach, where the latter means preserving Poincar\'e covariance, chiral and electromagnetic current-conservation, etc.  The Dyson-Schwinger equations (DSEs) provide such a framework.  Experimental and theoretical studies are underway that will identify observable signals of $M(p^2)$ and thereby explain the most important mass-generating mechanism for visible matter in the Standard Model.

This is an exciting time in hadron physics.  Through the DSEs, we are positioned to unify phenomena as apparently diverse as the: hadron spectrum; hadron elastic and transition form factors, from small- to large-$Q^2$; and parton distribution functions.\cite{Holt:2010vj}  The key is an understanding of both the fundamental origin of nuclear mass and the far-reaching consequences of the mechanism responsible; namely, DCSB.  These things might lead us to an explanation of confinement, the phenomenon that makes nonperturbative QCD the most interesting piece of the Standard Model.

\smallskip

%\section*{Acknowledgments}
\hspace*{-\parindent}\textbf{Acknowledgments}.
We acknowledge valuable discussions with C.~Hanhart, T.-S.\,H.~Lee, V.~Mokeev, S.~Riordan, S.\,M.~Schmidt, P.\,C.~Tandy and B.~Wojtsekhowski.
This work was supported by:
Forschungszentrum J\"ulich GmbH (HLLR, CDR);
the National Natural Science Foundation of China, contract no.\ 10705002 (LC);
the U.\,S.\ Department of Energy, Office of Nuclear Physics, contract nos.~DE-AC02-06CH11357 (HLLR, CDR) and DE-FG03-97ER4014 (ICC);
and the Department of Energy's Science Undergraduate Laboratory Internship programme (HLLR).
%
%and the U.\,S.\ National Science Foundation, under grant no.\ PHY-0903991, in conjunction with a CONACyT Mexico-USA collaboration grant.

%\bibliographystyle{ws-procs9x6}
%\bibliography{CDR10E}

\end{document}